\documentclass[10.5pt]{article}
\usepackage{graphicx}
\usepackage{amsmath,multirow}
\usepackage{enumerate,array}
\usepackage{amsfonts}
\usepackage{amssymb}
\usepackage{epsfig}
\usepackage{graphics}
\usepackage{multirow}
\usepackage{enumitem}
\usepackage{color}
\usepackage{lineno}
\usepackage{amscd}
\makeatletter
\renewcommand\@biblabel[1]{}
\makeatother

\renewcommand{\baselinestretch}{1.0}

\begin{document}
	
\title{ {Modeling climate extremes using the four-parameter \\ kappa distribution for $r$-largest order statistics}}

\author{Yire Shin,$^{1}$ Jeong-Soo Park$^{1,*}$  \\
	\small\it 1: Department of Mathematics and Statistics, Chonnam National University, Gwangju 61186, Korea\\
	\small\it *: Corresponding author, E-mail: jspark@jnu.ac.kr
}
\date{}
\footnotetext{Some parts of this paper were presented at the conferences of ECAS2021 and STAHY2021.}
\maketitle 

\begin{abstract}
Accurate estimation of the T-year return levels of climate extremes using statistical distribution is a critical step in the projection of future climate and in engineering design for disaster response. We show how the estimation of such quantities can be improved by fitting  {the four-parameter kappa distribution for $r$-largest order statistics} (rK4D), which was developed in this study. The rK4D is an extension of  {the generalized extreme value distribution for $r$-largest order statistics} (rGEVD), similar to the four-parameter kappa distribution (K4D), which is an extension of the generalized extreme value distribution (GEVD). This new distribution (rK4D) can be useful not only for fitting data when three parameters in the GEVD are not sufficient to capture the variability of the extreme observations, but also in reducing the estimation uncertainty by making use of the r-largest extreme observations instead of only the block maxima. We derive a joint probability density function (PDF) of rK4D and the marginal and conditional
cumulative distribution functions and PDFs. To estimate the parameters, the maximum likelihood estimation and the maximum penalized likelihood estimation methods were considered.
The usefulness and practical effectiveness of the rK4D are illustrated by the Monte Carlo simulation and by an application to the Bangkok extreme rainfall data.
A few new distributions for $r$-largest order statistics are also derived as special cases of the rK4D, such as the $r$-largest logistic, the $r$-largest generalized logistic, and the $r$-largest generalized Gumbel distributions. These distributions for $r$-largest order statistics would be useful in modeling extreme values for many research areas, including hydrology and climatology.
\textcolor{blue}{This paper was published at March, 2023: Shin, Y., Park, J-S. Modeling climate extremes using the four-parameter kappa distribution for $r$-largest order statistics, Weather and Climate Extremes, 39, 100533 (2023) https://doi.org/10.1016/j.wace.2022.100533. In this revision, some modification and correction from the published one are made on sentences and formula with blue color.}
\end{abstract}



\section{Introduction}
Severe weather damage to society has motivated many studies on extreme weather and climate events. Some of these studies have focused on the intensity and frequency of climate extremes.
Accurate estimation of quantities such as T-year return levels of climate extremes is a critical step in the projection of future climate and in engineering design for disaster response.
These quantities are usually estimated by applying a statistical distribution to extreme observations. 
One approach employs the generalized extreme value distribution (GEVD) to a series of block maxima.
Another method is to apply the generalized Pareto distribution to the peaks-over-threshold, in which the sample consists of observations above a specified threshold.

The GEVD has been widely used to analyze univariate extreme
values in many research areas, including hydrology and atmospheric sciences (Beirlant et al.~2004; AghaKouchak et al.~2013; Kharin et al.~2013; Naghettini 2017; Kim et al.~2020; Blanchet et al.~2021). 
The GEVD encompasses three possible asymptotic extreme value distributions derived by large sample theory. 
The probability density function (PDF) of the GEVD 
is as follows (Hosking and Wallis 1997):
\begin{linenomath*}\begin{equation} \label{pdf-gevd}
	f_3(x) = \sigma^{-1} \left(1 -k {{x-\mu} \over {\sigma}}\right) ^{{1 \over k} -1}\times F_3(x).
	\end{equation}\end{linenomath*}
where
\begin{linenomath*}\begin{equation} \label{cdf-gevd}
F_3(x) = \text{exp} \left\{ - \left(1 -k {{x-\mu} \over {\sigma}}\right) ^{1/k} \right\}.
\end{equation}\end{linenomath*}
is the cumulative distribution function (CDF) of the GEVD,
when $1-k (x- \mu ) / \sigma > 0$ and $\sigma>0$, where $\mu ,\; \sigma $, and $k$ are the
location, scale, and shape parameters, respectively. The particular case for
$k=0$ in Eq. (\ref{pdf-gevd}) is the Gumbel distribution. Note that the sign of $k$ in the above equations is changed from the book by Coles (2001).

One disadvantage of the GEVD is the limited amount (only the block maxima) of the data for model fitting. Because extreme values are scarce, making effective use of available information is important in extremes. This issue has motivated the search for a model that uses more data than the block maxima. The above univariate result
was extended to the $r$-largest order statistics model, which provides the joint density function of the limit
distribution (Coles, 2001);
\begin{linenomath*}\begin{equation} \label{rgevd}
f_3(x^{(1)},x^{(2)}, \cdots,x^{(r)}) = \sigma^{-1} \text{exp} \left\{ - w(x^{(r)})^{1/k} \right\}.
   \times \prod_{s=1}^r w(x^{(s)}) ^{{1 \over k} -1} ,
\end{equation}\end{linenomath*}
where $x^{(1)}\ge x^{(2)}\ge \cdots \ge x^{(r)}$, and
$ w(x^{(s)})= 1 -k {{x^{(s)}-\mu} \over {\sigma}} >0$ for $s=1,2,\cdots,r$.
This model is referred to as the rGEVD in this study.
The above PDF is not the same as that of the $r$-largest order statistics of a random sample of the GEVD, but it is the PDF of the limiting joint distribution of suitably re-scaled 
$r$-largest order statistics from any distribution (Smith, 1986).

The  {GEV models for $r$-largest order statistics} were developed by
Smith (1986) and Tawn (1988), building off of theoretical developments from Weissman (1978). 
The inclusion of more data up to the $r$-th order statistics, other than just one maxima in each block, will improve the precision of model estimation, while the interpretation of parameters is unchanged from the univariate GEVD.
The rGEVD was encouraged by Zhang et al.~(2004) and has been employed in many real applications (An and Pandey 2007; Wang and Zhang 2008; Feng and Jiang 2015; Naseef and Kumar 2017).
The number $r$ comprises a bias-variance trade-off: small values of $r$ consist of few data points leading to high variance, whereas large values of $r$ are likely to violate the asymptotic support for the model, leading to a bias (Coles, 2001). Bader et al.~(2017) and Silva et al.~(2021) developed automated methods for selecting $r$ from the rGEVD.

The distributions derived by the extreme value theory are not always applicable to real extreme observations.
For small to moderate sample sizes, GEVD sometimes yields inadequate results. This may be because GEVD is derived from a large sample theory for the extremes of independent sequences.
Similarly, a variety of studies have shown that the GEVD may not represent the observed skew-kurtosis relationship of extreme observations (Hosking, 1994; Salinas et al.~2014; Nathan et al.~2016; Kjeldsen et al.~2017). For example, the log-Pearson type III or the generalized logistic distributions have been used to model extreme hydrological events, and their use is sometimes mandated by government agencies (Vogel and Wilson 1996).

As a generalization of some common three-parameter distributions including the GEVD, the four-parameter kappa distribution (K4D) was introduced by Hosking (1994).
The K4D can be useful to represent a wider range of skew-kurtosis pairings and capture more variability of observations than the three parameter distributions.
The K4D has been employed for regional frequency analysis (Hosking and Wallis 1997), drought frequency analysis (Núñez et al.~2011), wet-day precipitation modeling (Ye et al.~2018), and rainfall bias correction for climate models (Heo et al.~2019). Some studies on the parameter estimation of the K4D have been conducted (Dupuis 1997; Dupuis and Winchester 2001; Singh and Deng 2003; Park and Kim 2007; Murshed et al.~2014; Seenoi et al.~2020; Papukdee et al.~2021).

The PDF of K4D is, for $ k\neq0,\; h\neq0 $, $\sigma>0$,
\begin{linenomath*}\begin{equation} \label{k4d}
	f_4(x) =\sigma^{-1}w(x)^{(1/k)-1} F_4(x)^{1-h}  ,
	\end{equation}\end{linenomath*}
where
\begin{linenomath*}\begin{equation} \label{w(x)}
	w(x) = 1 -k {{x-\mu} \over {\sigma}},
	\end{equation}\end{linenomath*} and
\begin{linenomath*}\begin{equation} \label{cdf4}
	F_4(x) =  {\left\{1-h\; {w(x)}^{1/k} \right\}}^{1/h},
	\end{equation}\end{linenomath*}
is the CDF of the K4D (Hosking, 1994).
 Note that a new shape parameter $h$ is added from the GEVD.  
 Subscripts 3 and 4 in $f$ or $F$ are hereafter used to indicate the corresponding functions of the (3-parameter) GEVD and the K4D, respectively.
 The first part of the PDF of K4D ($\sigma^{-1}w(x)^{(1/k)-1}$) is the same as the first part of the PDF of the GEVD in Eq. (\ref{pdf-gevd}). The second part $F_4 (x)^{1-h}$ in K4D is an `unlimited' form of $F_3 (x)$ in GEVD, in the sense that $\lim_{h \to 0} F_4 (x)^{1-h} = F_3 (x)$. For the extension of rGEVD to  {the K4D for $r$-largest order statistics} in the next section, we follow the same method.

 \begin{figure}[!t]
 	\centering
 	\begin{tabular}{l}	\includegraphics[width=13cm, height=6.cm]{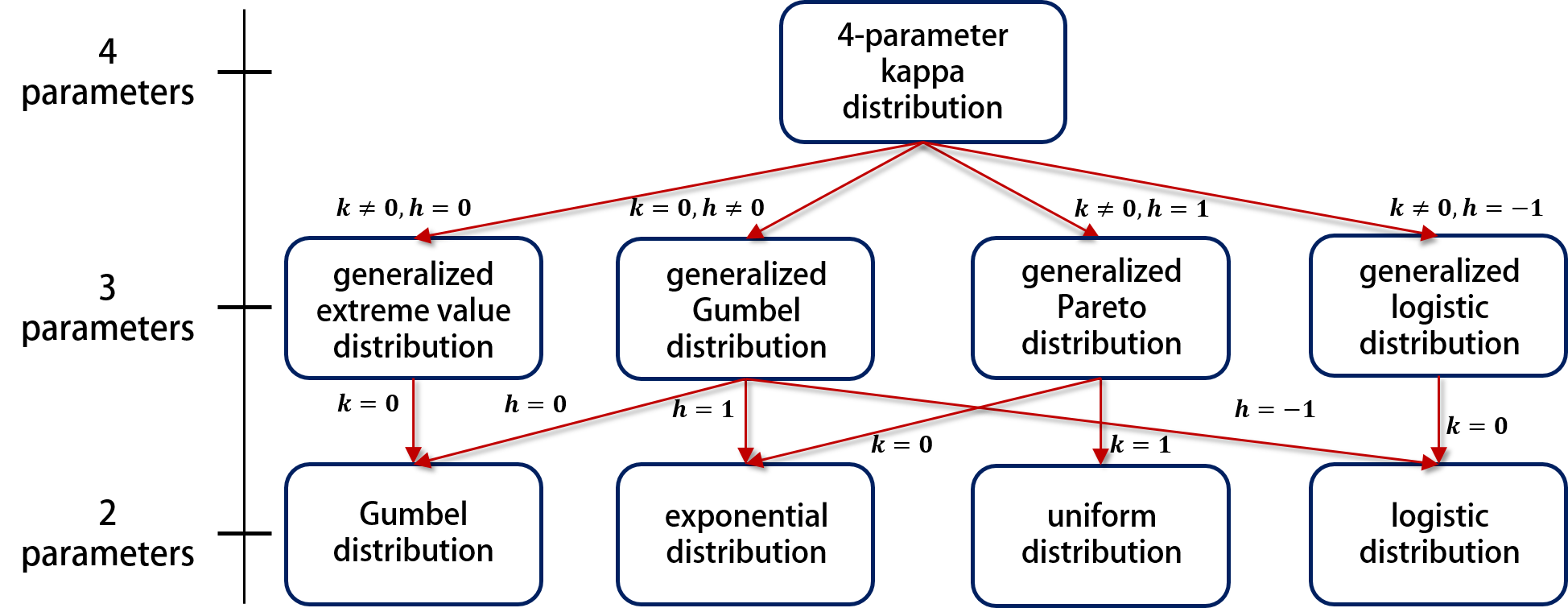} \end{tabular}	
 	\caption{Relationship of the four-parameter kappa distribution (K4D) to other distributions, which indicates a wide coverage of K4D.} \label{K4D_rel}
 \end{figure}

The K4D includes many distributions as special cases, as shown in
Figure \ref{K4D_rel}: the generalized logistic distribution for $h=-1$, the GEVD for
$h=0$, the generalized Pareto distribution for $h=1$, the generalized Gumbel distribution for $k=0$, and the Gumbel distribution for $h=0,~k=0$.
The K4D is widely applicable to data including extreme values as well as skewed data. It has been applied in many fields, particularly in the hydrology and atmospheric sciences (e.g., Parida 1999; Park and Jung 2002; Seo et al.~2015; Blum et al.~2017; Kjeldsen et al.~2017; Jung and Schindler 2019).

\begin{figure}[!t]
 	\centering
 	\begin{tabular}{l}	\includegraphics[scale = 0.35]{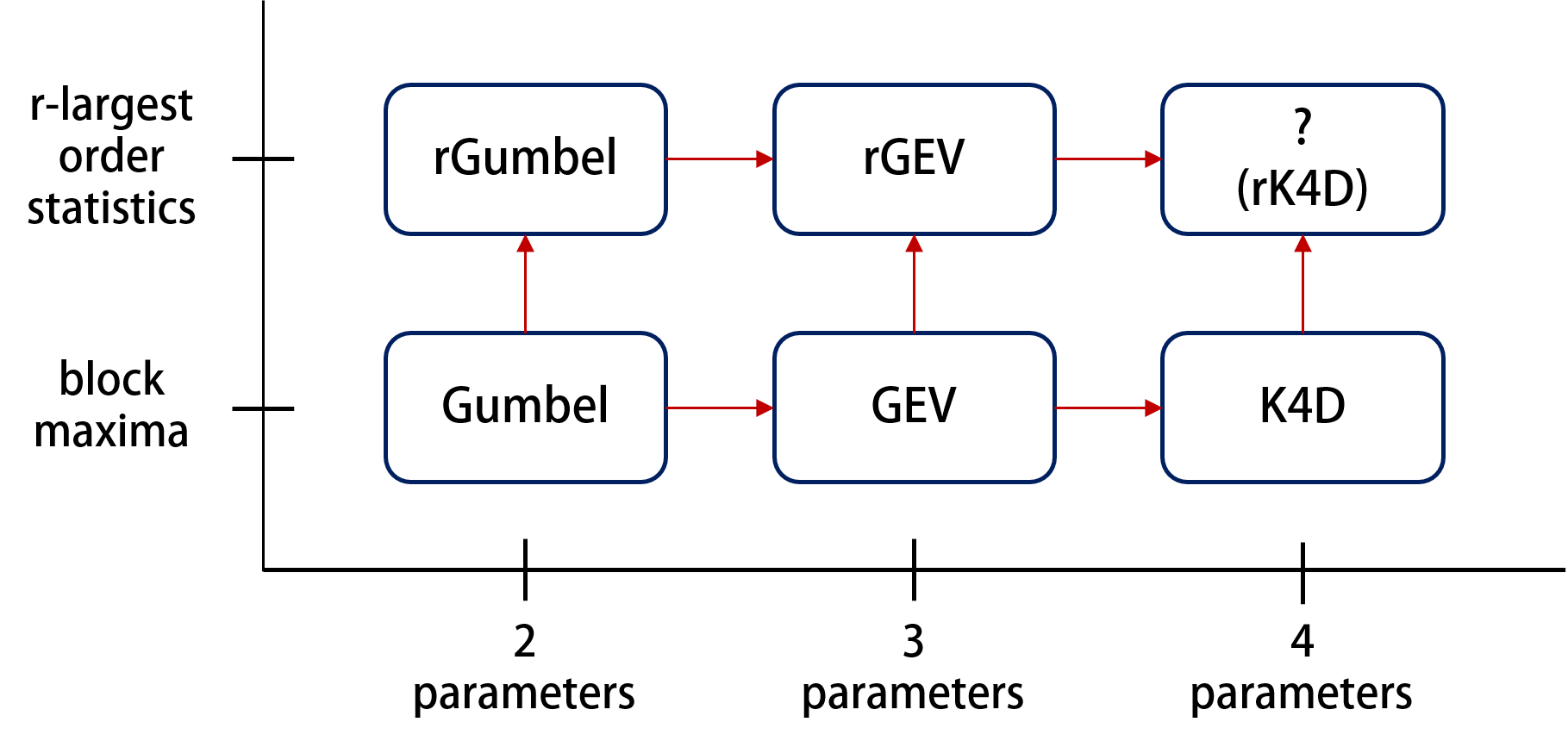} \end{tabular}	
 	\caption{A motivation schema on generalizations from 2 parameters to 4 parameters, and extensions from the block maxima models to
 		the $r$-largest extremes model, which leads to  {four-parameter kappa distribution for the $r$-largest order statistics} (rK4D).} \label{Fig:motiv}
 \end{figure}

In analyzing extreme values, K4D has the same limitation of using only the block maxima as the GEVD has.
As with GEVD extended to rGEVD, an extension of the K4D to the $r$-largest order statistics model may be useful in addressing
this limitation. The inclusion of more observations up to $r$-th order statistics
other than just one maxima per block will improve the precision of model fitting.
In this study, we thus developed an $r$-largest extreme model as an extension of the K4D and rGEVD.
This is referred to as the rK4D in this study. Figure \ref{Fig:motiv} illustrates the motivation schema.

The remainder of this paper is organized as follows. Section 2 defines the rK4D. The maximum and penalized likelihood estimations of the parameters are discussed in Section 3.
 Sections 4 and 5 illustrate the usefulness and practical effectiveness of rK4D by Monte Carlo simulation and by application to the Bangkok extreme rainfall data.
 In Section 6, some new $r$-largest distributions as special cases of rK4D are derived.
 The discussion is presented in Section 7, followed by a conclusion in Section 8. The Appendix includes the details of marginal and conditional distributions, random number generations, the delta method for variance estimation, and the profile likelihood approach for confidence interval.

\section{r-largest four parameter kappa distribution}
\subsection{Definition of the rK4D}

Let us denote $\underline{x}^r = (x^{(1)},x^{(2)},\cdots,x^{(r)})$ as the $r$-largest order statistics from any distribution.
 {The four-parameter kappa distribution for $r$-largest order statistics} (rK4D) is an analogous extension from the rGEVD by following a similar method as the extension from the GEVD to the K4D.

We define the joint PDF of the rK4D; under $k\neq0,\; h\neq0$,
\begin{linenomath*}\begin{equation} \label{rk4d}
f_4 (\underline{x}^r) = \sigma^{-r} C_r \times g(\underline{x}^r) \times F_4 (x^{(r)} )^{1-rh}.
\end{equation}\end{linenomath*}
where
\begin{linenomath*} \begin{equation}  \label{C_r}
	C_r = 	\left\{ \begin{array}{ll} & \prod_{i=1}^{r-1}\; [1-(r-i)h], ~~~\textrm{if} ~ r\ge 2,  \\
	& 1, ~~~~~~~~~~~~~ \textrm{if}~ r=1, \end{array} \right. 	
	\end{equation}\end{linenomath*}	
 $F_4$ is the cumulative distribution function (CDF) of K4D, as shown in Eq.~(\ref{cdf4}), and
\begin{linenomath*}\begin{equation} \label{g(x)}
g(\underline{x}^r) = \prod_{s=1}^r w(x^{(s)})^{{1 \over k} -1} ,
\end{equation}\end{linenomath*}
where $w(x)$ is defined in Eq. (\ref{w(x)}); $ w(x^{(s)})= 1 -k {{x^{(s)}-\mu} \over {\sigma}}$. From (\ref{C_r}), we have the condition that $h < {1 \over {r-1}}$ for $r \ge 2$.

The support of this PDF is $x^{(1)}\ge x^{(2)}\ge \cdots \ge x^{(r)}$, $\sigma>0$,
$ w(x^{(s)}) >0$ for $s=1,2,\cdots,r$, $C_r > 0$,
and $ 0< 1-h\; w(x^{(r)})^{1/k} <1$. 	
When $r=1$, the PDF is the same as the PDF of K4D in Eq. (\ref{k4d}). When $h \rightarrow 0$,
this PDF goes to the rGEVD PDF in Eq.~(\ref{rgevd}).

Note that $\sigma^{-r} \times g(\underline{x}^r)$ in the above PDF is the same as the quantity in the rGEVD PDF. The term $F_4 (x^{(r)} )^{1-rh}$ in the above PDF is an `unlimited' form of $F_3 (x^{(r)} )$ in the PDF of rGEVD. The constant term $C_r$ is included to make $f_4 (\underline{x}^r)$ become a PDF, in the sense that $\int\int \cdots \int f_4 (\underline{x}^r) d\underline{x}^r$ becomes 1.

Note again that the PDF of rK4D defined in this study is not the same as the PDF of the $r$-largest order statistics of a random sample from the K4D. The PDF of rK4D is an 'unlimited' extension of the PDF of the limiting joint distribution of suitably re-scaled 
$r$-largest order statistics from any distribution.

Appendix A contains the sections on the consistency of the definition of the rK4D, the marginal PDF and CDF of the $s$-th order statistics from the rK4D, the conditional distributions of rK4D, and the random number generation from rK4D. We see that these quantities derived from rK4D go to the corresponding quantities of rGEVD, as $h \rightarrow 0$.

\section{Parameter estimation}

\subsection{Maximum likelihood estimation}

Let $\underline{x}^r_i =((x^{(1)}_i, x^{(2)}_i,\cdots,x^{(r)}_i)$, which is the $i$-th
observation of the $r$-largest order statistics for $i=1,2,\cdots,m$, where $m$ is the sample size.
By assuming $\{\underline{x}^r_1, \underline{x}^r_2, \cdots, \underline{x}^r_m \}$ follow the rK4D,
the likelihood function of $\mu, \sigma, h, k$ is as follows for $k\neq0,\; h\neq0$;
\begin{linenomath*}\begin{equation} \label{lfrk4d}
	L(\mu, \sigma, h,  k | \underline{x}^r ) = \prod_{i=1}^m \left[ \sigma^{-r} C_r F_4 (x^{(r)}_i)^{1-rh}.
	\prod_{j=1}^r \left( 1- k \frac{(x^{(j)}_i -\mu)}{\sigma} \right) ^{{1 \over k}-1} \right],
\end{equation}\end{linenomath*}	
under constraints. The details of the constraints (Hosking and Wallis, 1997), which should be specified in minimizing numerically
the negative log-likelihood function with respect to the parameters is $\sigma >0$, $ h < {1 / (r-1)}$ for $r \ge 2$,
\begin{equation}
	 h < \textrm{min}_{\substack{1 \le i \le m}} \; \left[ {1 \over {w(x^{(r)}_i)^{1/k}} } \right].
\end{equation}
	 and
\begin{linenomath*}\begin{equation} \begin{cases}
\textrm{max}_{\substack{1 \le i \le m}} \;w(x^{(1)}_i) < {\sigma \over k } + \mu, &~\textrm {if} ~k > 0\\
\textrm{min}_{\substack{1 \le i \le m}} \;w(x^{(r)}_i) > {\sigma \over k } + \mu, &~\textrm {if} ~k < 0.
\end{cases}
\end{equation}\end{linenomath*}	
We implemented a numerical algorithm using the `Rsolnp' package in R program (Ghalanos et al.~2012).

The standard errors of the maximum likelihood estimates (MLE) were obtained approximately using the squared root
of the diagonal terms of the inverse of the observed Fisher information matrix.
 {An accurate computation when $k \rightarrow 0$ in rK4D was achieved
by using a R function `expm1' as mentioned in Bader et al.~(2017). See the Supplementary Material for more details.}

\subsection{Maximum penalized likelihood estimation}

It is known that the MLE of K4D sometimes does not perform well. To address this problem,  Papukdee et al.~(2021)
proposed the maximum penalized likelihood estimator (MPLE) for K4D, which performed better than MLE. Thus, we consider the MPLE of the parameters for the rK4D in this study. The MPLE in rK4D was obtained by minimizing 
the penalized negative log likelihood function which is
\begin{equation}
l_\text{pen}(\mu,\sigma,k,h) = - \textrm{ln}(L(\mu,\sigma,k, h)) - \textrm{ln}(p(k,h)), \label{MPLE1}
\end{equation}
where $ p(k,h) $ is the joint penalty function (PF) of the two shape parameters $ k $ and $ h $. In this study, we treat $ k $ and $h$ as independent, so that $ p(k,h) = p(k)\times p(h) $.

Among the many penalty functions, we employ Coles and Dixon's PF for $k$ and the adjusted Martins and Stedinger's PF for $h$ (Martins and Stedinger 2000). The Coles and Dixon PF for $k$ are 
\begin{equation}\label{p_coles}
p(k) = \begin{cases} 1 & \mbox{if $k \geq 0$}\\
\text{exp}\{-\lambda(\frac{1}{1+k}-1)^{\alpha}\} & \mbox{if $-1< k < 0$}\\
0 & \mbox{if $k\leq -1$}\end{cases}
\end{equation}
for non-negative values of $\alpha$ and $\lambda$. For hyperparameters $\alpha$ and $\lambda$,
Coles and Dixon (1999) suggested using a combination of $\alpha=1$ and $\lambda=1$.

 {Following Papukdee et al.~(2022), we chose an adjusted penalty function of Martins and Stedinger (2002) for $h$.
    The adjusted penalty function is a beta function with parameters 6 and 9 defined over the range (-1.2, 1.2). The beta shape parameters 6 and 9 are same with
    those in Martins and Stedinger (2002). The range of $h$ was adopted from Dupuis and Winchester (2001).}
     { In the rK4D case, however, we have another constraint that $h<1/(r-1)$.
   Thus the adjusted Martins and Stedinger PF for $h \in (-1.2,\; b)$ where $b=1.2$ for $r=1$ and $b=1/(r-1)$ for $r > 1$, in the rK4D, is
	\begin{eqnarray}
		p(h) & = &  \dfrac{(1.2+h)^{u-1}(b -h)^{v-1}}{B_E (u,v)} \label{penms}
	\end{eqnarray}
	with $u=6$ and $v=9$ (Martins and Stedinger, 2002; Papukdee et al.~2021), where
	\begin{eqnarray*}
		B_E (u,v) &=& \int_{-1.2}^{b} (1.2+h)^{u-1}(b-h)^{v-1} dh
	\end{eqnarray*}
	is the modified beta function.}
	
	Appendix B contains the sections on the quantiles of the $s$-th order statistic of the rK4D, the delta method for variance estimation, and the profile likelihood for the confidence interval of the return level in the rK4D.

\section{Monte Carlo simulation}

We conducted a Monte Carlo simulation to investigate the performance of rK4D on heavy-quantile estimation ($1-p=.99$).  A total of 1000 samples were used to obtain the bias, variance, and mean squared error (MSE) of the 100-year return level of the maximum. The bias, variance, and MSE were defined as
\begin{eqnarray} \label{BIAS} 
\text{Bias }    & = & {q- \bar{\hat{q}}} , ~~~
\text{Var} =  \frac{1}{M}\sum^{M}_{j=1}  (\hat{q}_{j}-\bar{\hat{q}} )^2,  \\
\text{MSE}  & = &  \text{Bias}^2 + \text{Var},   
\end{eqnarray}
where $\hat{q}_{j}$ and $q$ are the estimated and true quantiles of the maximum, respectively; $\bar{\hat{q}}= \frac{1}{M}\sum^{M}_{j=1} \hat{q}_{j}$, and $M$ is the number of successful convergence among 1000 trials. 

\begin{table}[h]
\centering	\caption{ {Six settings of parameters for Monte Carlo experiments in the $r$-largest four parameter kappa distribution}}
	\label{paraset}
	\begin{tabular}{c|c|c|c|c} \hline
Experiment & $\mu$  & $\sigma$ & $k$ & $h$   \\ \hline
1: & \multirow{6}{*}{100} & \multirow{6}{*}{10} & \multirow{3}{*}{-0.3} & -0.1 \\  \cline{5-5}
2: & & & & -0.3 \\  \cline{5-5}
3: & & & & 0.1  \\  \cline{4-5}
4: & & & \multirow{3}{*}{0.1}  & -0.1 \\  \cline{5-5}
5: & & & & -0.3 \\ \cline{5-5}
6: & & & & 0.1 \\ \hline
\end{tabular}
\end{table}

{In generating random number from the rK4D, the property in Appendix A.4 was used. The property exploited here is that Eq.~(\ref{condcdfk4d}) is \textcolor{blue}{the $({1-(s-1)h})$ power of } the CDF of the K4D with right truncated at $x^{(s-1)}$. Similarly to the rGEVD case presented by Bader et al.~(2017), we sampled random numbers from a truncated
	uniform distribution and used the quantile function Eq.~(\ref{z_pk4d}) to transform it to generate random numbers of rK4D. Detailed algorithm is provided in {the Appendix A.5.} 

\begin{figure}[ht!]
	\centering
	\begin{tabular}{l}	\includegraphics[width=15.5cm, height=12.5cm]{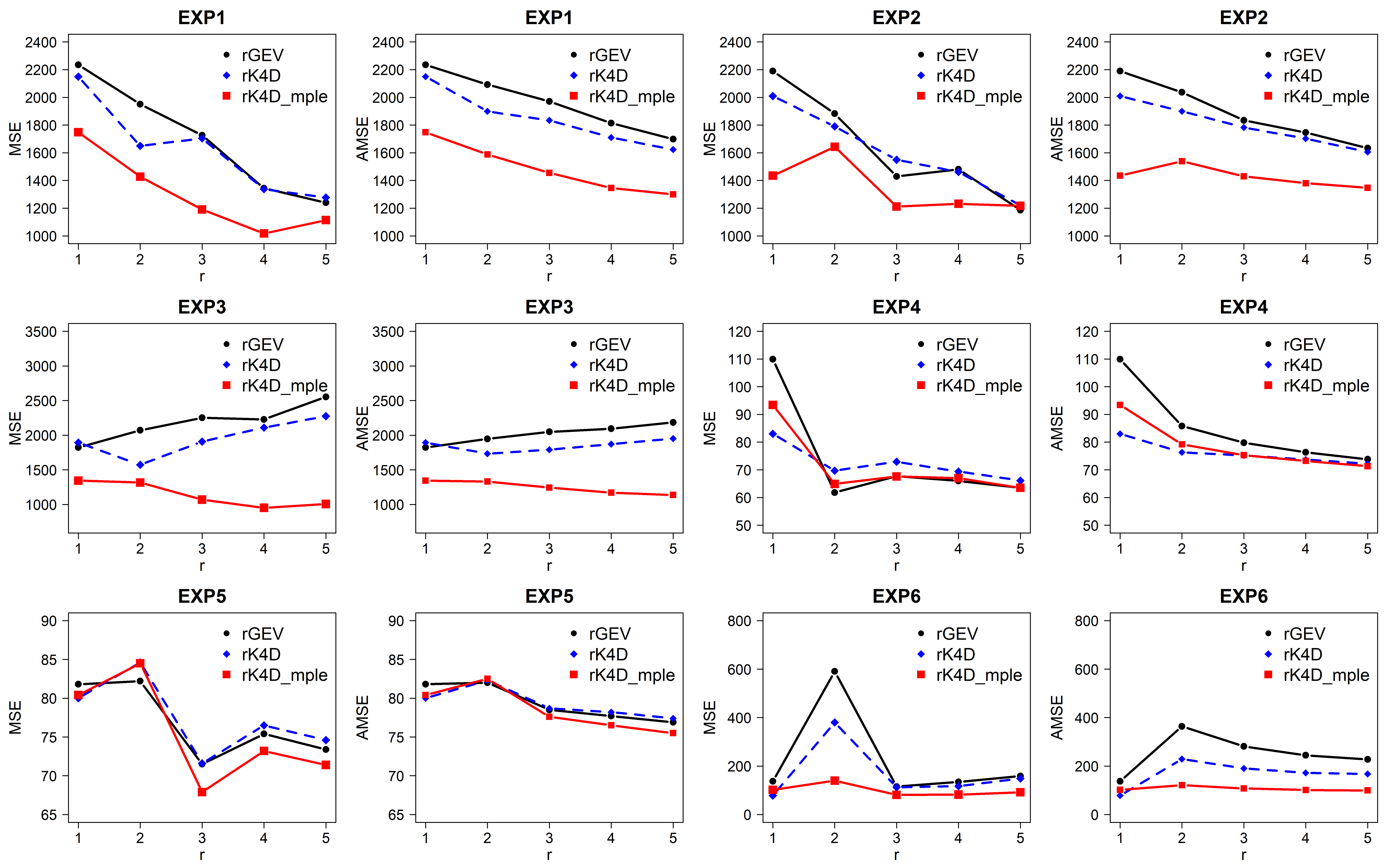} \end{tabular}	
	\caption{Result of six Monte Carlo experiments from EXP1 to EXP6: the mean squared error (MSE) and the averaged mean squared error (AMSE) for 100-year return level estimation as $r$ changes from 1 to 5 for three different methods. For rK4D, the maximum likelihood estimator (MLE) and maximum penalized likelihood estimator (MPLE) are employed. MLE is used for the rGEVD. } \label{MC-MSE}
\end{figure}

To evaluate the estimation performance for the $s$-th order statistics, where $s=1,2,\cdots, r$, we consider the MSE of the 100-year return level for the $s$-th order statistic.  {We denote this by $\text{MSE}_{s}$, where $\hat{q}_{j}$ and $q$ in Eq. (\ref{BIAS}) are the estimated and true quantiles of the $s$-th order statistics. Then, our new measure to evaluate the total estimation ability is the averaged MSE up to $r$, which is defined by
\begin{equation} \label{AMSE}
\text{AMSE}(r) = (\text{MSE}_{s=1} + \text{MSE}_{s=2} + \cdots + \text{MSE}_{s=r})/r.
\end{equation}
}
We performed Monte Carlo experiments for six combinations of parameters (see Table \ref{paraset}) with $r=1,2,\cdots,5$. The parameters $\mu$ and $\sigma$ are set to 100 and 10, respectively, because they are equivariant. For the parameter estimation in rK4D, MLE and MPLE are employed. For comparison, rGEVD with MLE was also considered.

Figure \ref{MC-MSE} shows the result of this simulation where the MSE and the AMSE are plotted for each $r$ and for each experiment. The detailed values including the bias and variance are provided in Tables S3 to S8 in the Supplementary Material. From these figures and tables, we can say that the rK4D with the MPLE works better than the rGEVD and the rK4D with the MLE. The performance of rGEVD comes close to that of rK4D with the MLE. It seems the rK4D with the MLE works the same as or a little better than the rGEVD.

     {It is notable that, for the cases when $k$ is positive and $h$ is negative, values of MSE and AMSE are relatively small compared to the cases of $k$ negative.
    The performance of three methods are similar in these cases (EXP4 and EXP5).
     But the estimation methods when $k$ is positive and $h$ is negative are not our primary interest.
     When both $k$ and $h$ are positive, values of MSE and AMSE are moderately large. In this case (EXP6), rK4D with MLE/MPLE works better than rGEV.}

      {The values of MSE and AMSE show decreasing pattern as $r$ increases from 1 to 5, except in the third experiment (EXP3).  It is expected that small values of $r$ generate few data leading to high variance while large values of $r$ are likely to violate the asymptotic support for the model, leading to bias (Coles, 2001). Thus,
          We think the optimal value of $r$ can be selected by a bias-variance trade-off; at the point when values of MSE begin to increase after continuous decrease as $r$ increases from 1.}

           {In the EXP3 where $k$ is negative and $h$ is positive, values of MSE and AMSE obtained from rK4D with MPLE are decreasing as $r$ increases. As seen in the EXP6, values of MSE and AMSE decrease slowly or increase. Thus we can say, from EXP3 and EXP6, that the benefit of using $r$ largest order statistics over the block-maxima model does not seem significant when $h$ is positive.}

\subsection{Testing hypothesis for $h$}

	 {A test procedure for $H_0: h = 0$ would be useful for practitioners to choose between
	rK4D and rGEV. For this purpose, a likelihood ratio test (LRT) is considered here
with $H_0: h = 0$ vs. $H_1: h \ne 0$.
We did a small Monte Carlo simulation experiment for for $r=2,3,4$ fixed with sample size $n=100$.
 The size ($\alpha=0.05$) and power of this test were obtained for $k=0.2$ using 1000 repetition for various $h$ values.}

\begin{figure}[h!]
\caption{ {The size ($\alpha=0.05$) and power of the likelihood ratio test for $k=0.2$ using 1000 repetition for various $h$ values in the four-parameter kappa distribution for $r$-largest order statistics.}} \label{power-k}
		\centering
		\begin{tabular}{l}	\includegraphics[width=11.5cm, height=6cm]{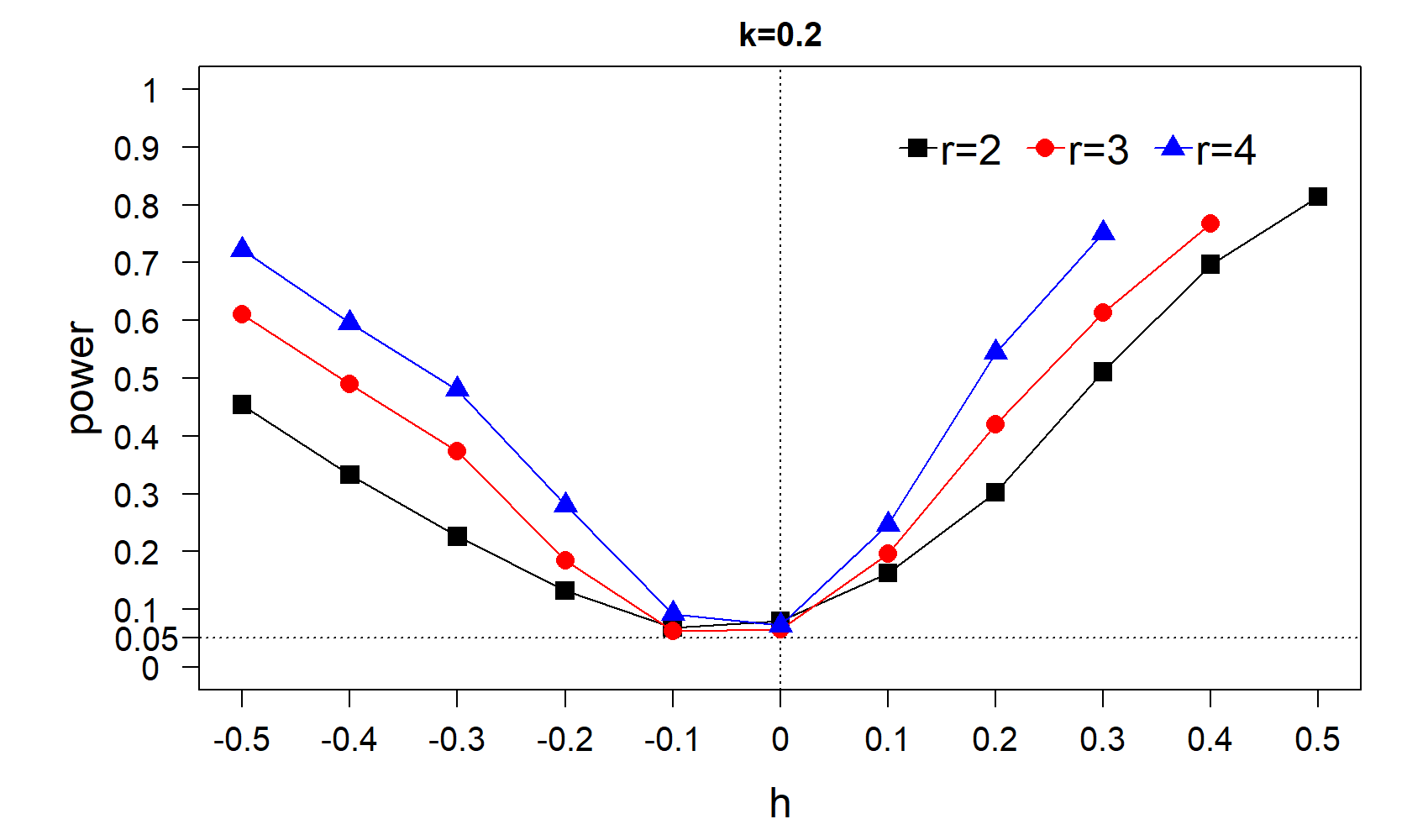} \end{tabular}	
\end{figure}
 
  {Figure \ref{power-k} and Table S2 in the Supplementary Material show the size and power obtained from our experiment. It seems that the size is well maintained. The power for $h$ positive looks better than the power for $h$ negative. It is notable that the power increases more rapidly as $r$ changes from 2 to 4. When $k$ is negative and $h$ is negative, the power turned out to be not good (not presented here).
 The study on the testing hypothesis to choose between rK4D and rGEV is out of scope of this paper but a topic of the next research.}

\section{Real application: Bangkok extreme rainfall data}

The top five annual daily rainfall events were taken from the daily records of a rain gauge station in Bangkok, Thailand, from to 1980-2018. This data is provided in Supplementary Material.
The rK4D model was fitted to the data for $r=1,2,\cdots,5$. The MLE and MPLE of the parameters and the 50-year return levels with standard errors in parentheses for several values of $r$ are given in Table \ref{tab:Bangkokdata}.
For comparison, similar results from the fitted rGEVD are presented in Table \ref{tab:Bangkokdata}. The upper table is for the rGEVD, the middle table is for the rK4D with the MLE, and the bottom table shows the rK4D with the MPLE.

\begin{table}[!t]
	\centering \caption{The estimates of parameters and 50-year return level ($r_{50}$) with standard errors (se) of the estimates in parenthesis which are obtained from the $r$-largest extremes models fitted to Bangkok rainfall data with different values of $r$.
		  The upper table is for the rGEVD, the middle table is for the rK4D/MLE, and the bottom table shows the rK4D/MPLE. `nllh' stands for the negative log-likelihood function value. \label{tab:Bangkokdata}} \vspace{.5cm}
\begin{tabular}{|c|c|c|c|c||c|}
	\hline
	r &  nllh & $\hat{\mu}$ (se)  & $\hat{\sigma}$ (se) & $\hat{k}$ (se)  & rGEVD $r_{50}$ (se) \\ \hline
	1 & 195.8  & 94.2 (5.3) & 28.3 (4.2)  & 0.166 (0.152) & 249.8 (50.8) \\ \hline
2 & 346.4  & 90.3 (4.3) & 27.5 (3.5)  & 0.266 (0.140) & 279.1 (70.8) \\ \hline
3 & 472.0  & 89.9 (4.0) & 28.1 (3.5)  & 0.276 (0.112) & 287.2 (67.8) \\ \hline
4 & 580.7  & 91.4 (3.8) & 28.0 (3.1)  & 0.206 (0.082) & 259.0 (46.3) \\ \hline
5 & 678.7  & 92.0 (3.7) & 27.8 (3.0)  & 0.188 (0.073) & 252.1 (41.6) \\ \hline

\end{tabular}

\vspace{0.5cm}
\begin{tabular}{|c|c|c|c|c|c||c|}
	\hline
	r & nllh &  $\hat{\mu}$ (se)  & $\hat{\sigma}$ (se) & $\hat{k}$ (se)   & $\hat{h}$ (se) & rK4D $r_{50}$ (se)  \\ \hline
	1 & 196.2  & 101.0 (4.8) & 22.6 (3.8) & -0.328 (0.145)  & -0.621 (0.144)  & 279.0 (69.2) \\
2 & 344.3  & 89.6 (4.1)  & 35.3 (5.7) & -0.019 (0.154)  & 0.340 (0.098)   & 232.8 (37.7)  \\
3 & 470.8  & 91.1 (3.7)  & 30.2 (3.6) & -0.147 (0.117)  & 0.157 (0.066)   & 250.1 (46.0)  \\
4 & 580.7  & 91.5 (3.7)  & 28.0 (3.1) & -0.200 (0.098)  & 0.009 (0.074)   & 257.2 (47.5)  \\
5 & 678.7  & 91.8 (3.9)  & 27.7 (3.1) & -0.199 (0.084)  & -0.018 (0.067)  & 255.3 (44.8)  \\
\hline
\end{tabular}

\vspace{0.5cm}
\begin{tabular}{|c|c|c|c|c|c||c|}
	\hline
	r & nllh &  $\hat{\mu}$ (se)  & $\hat{\sigma}$ (se) & $\hat{k}$ (se)   & $\hat{h}$ (se) & MPLE $r_{50}$ (se)  \\ \hline
	1& 196.0 & 95.9 (6.9) & 27.0 (6.5)  & -0.170 (0.182)  & -0.104 (0.381)   & 245.4 (47.7) \\
2& 345.7 & 89.9 (3.9) & 33.1 (5.6)  & -0.058 (0.158)  & 0.289 (0.127)    & 235.1 (38.7) \\
3& 471.5 & 91.0 (3.7) & 29.7 (3.3)  & -0.148 (0.124)  & 0.144 (0.074)    & 247.9 (46.1) \\
4& 581.0 & 91.4 (3.9) & 27.7 (3.1)  & -0.196 (0.093)  & 0.000 (0.078)    & 253.7 (44.8) \\
5& 678.9 & 91.6 (3.5) & 27.5 (2.8)  & -0.195 (0.081)  & -0.026 (0.077)   & 252.0 (41.1) \\
\hline
\end{tabular}
\end{table}

In Table \ref{tab:Bangkokdata}, it is notable that the estimates of the 50-year return levels ($r_{50}$) are close to a value around 253 as $r$ rises to 5 for all three methods. The standard errors of the 50-year return level are also close to a value of approximately 42 as $r$ rises to five for all three methods. This indicates that the estimation becomes stable as $r$ rises to a certain value, such as 5, regardless of the method.
The standard error of the 50-year return level was calculated using the delta method (Coles, 2001). In most cases of $r$, the standard errors obtained by rK4D with MPLE are lower than those obtained by rGEVD and rK4D with MLE. 

 {For choosing between rK4D and rGEV, we carried the LRT for this data. As seen in Table \ref{tab:Bangkokdata}, the null hypothesis ($H_0: h=0$) was rejected only for $r=2$. If we apply the Wald test for $h$ based on
     $\hat h $ and the standard error of $\hat h $, null hypothesis is rejected for $r=1,2$, and 3, whereas it is accepted for $r=4$ and 5.}

\begin{figure}[!h]
	\centering
	\begin{tabular}{l}	\includegraphics[width=15cm, height=13cm]{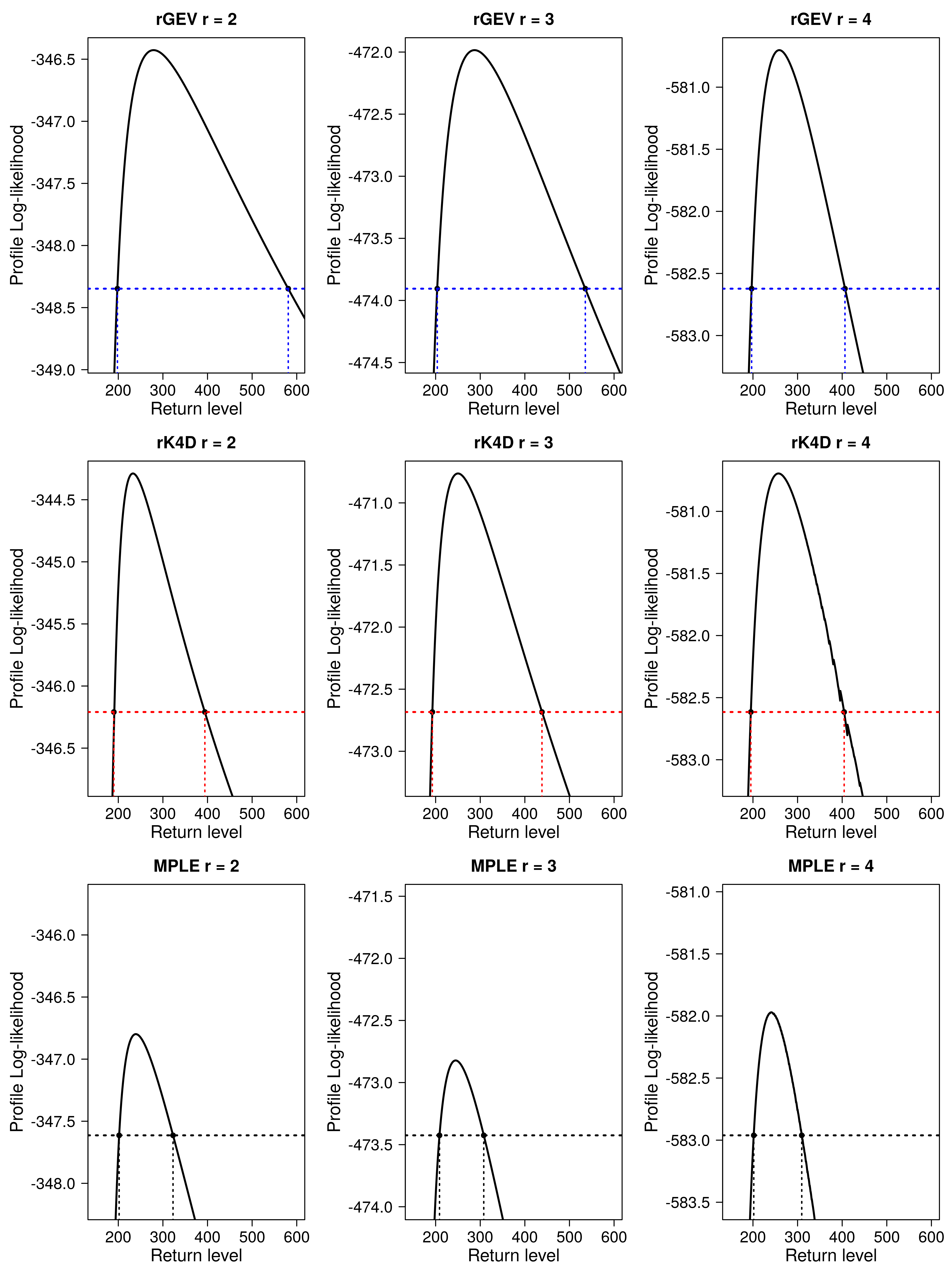} \end{tabular}	
	\caption{The 95\% profile likelihood confidence intervals for the 50-year return levels estimated by three different methods for $r$ = 2, 3, and 4. MLE is used for rGEVD, while MLE and MPLE are used for rK4D. The top panel shows the results from the rGEVD, and the middle and bottom panels are the results from the rK4D/MLE and rK4D/MPLE, respectively. The confidence intervals from the rK4D/MPLE are the smallest.} \label{pro_rl}
\end{figure}

For the cases when $r$ equals 1 or 2, it appears that the return levels and standard errors are unstable for rGEVD and rK4D/MLE. However, as $r$ rises to 4 or 5, these values become stable.
This phenomenon was also observed in the Monte Carlo simulation study. 
Thus, we would suggest that at least $r=3$ or $r=4$ are recommended for use of the rGEVD or rK4D, although we need to select an appropriate value of r for real applications.

Figure \ref{pro_rl} shows the 95\% profile likelihood confidence intervals for the 50-year return levels estimated by three different methods for $r$ = 2, 3, and 4. Top panel is results from the rGEVD/MLE, the middle and bottom panels are results from the rK4D/MLE and the rK4D/MPLE, respectively. The confidence interval lengths, estimated by the rK4D/MPLE, are the smallest among the three methods while those intervals estimated by rGEVD are the widest.

Based on the return level estimation and confidence interval study for the Bangkok extreme rainfall data and from the Monte Carlo simulation study presented in the previous section, we can conclude that the rK4D with the MPLE works well compared to the rGEVD and rK4D with the MLE.


\begin{figure}[!h]
	\centering
	\begin{tabular}{l}	\includegraphics[width=14cm, height=9.cm]{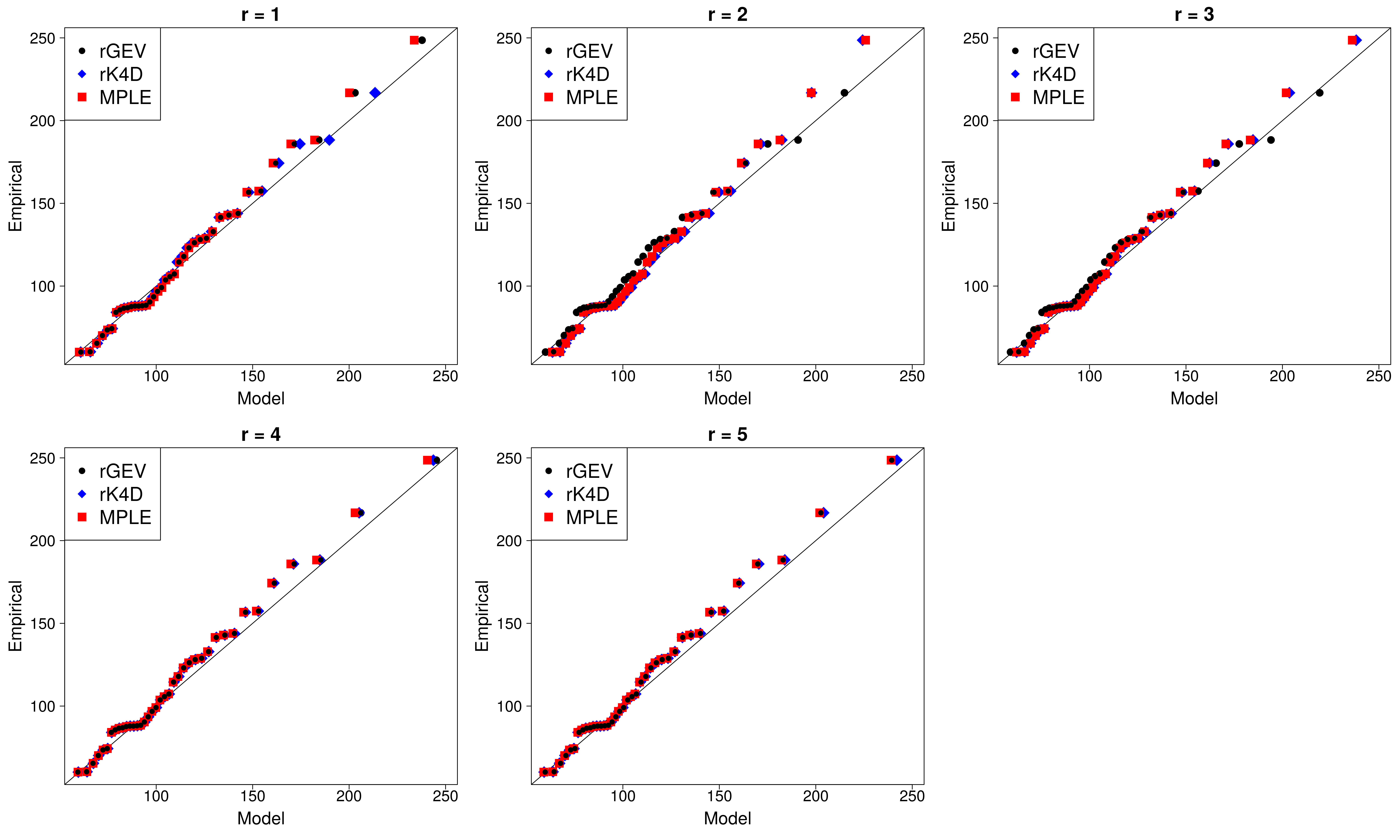} \end{tabular}	
	\caption{Quantile-per-quantile plots obtained from the largest order statistics for the rK4D/MLE fit (blue points), for the rGEVD fit (black points), and for the rK4D/MPLE (red points) to  the Bangkok extreme rainfall data with several values of $r=1,2,\cdots,5$.} \label{QQplot-1}
\end{figure}

Figure \ref{QQplot-1} shows quantile-per-quantile plots  obtained from the largest ($s=1$) order statistics for the rK4D/MLE fit (blue points), for the rGEVD fit (black points), and for the rK4D/MPLE (red points) to  Bangkok rainfall data with several values of $r$. In this figure, we see that the rK4D/MLE works well for $r=1$ while the rGEVD works well as $r$ equals 2 or 3. For the cases when $r$ equals 4 or 5, there seems no difference among the three methods.  



\begin{figure}[!h]
	\centering
	\begin{tabular}{l}	\includegraphics[width=14.5cm, height=8.cm]{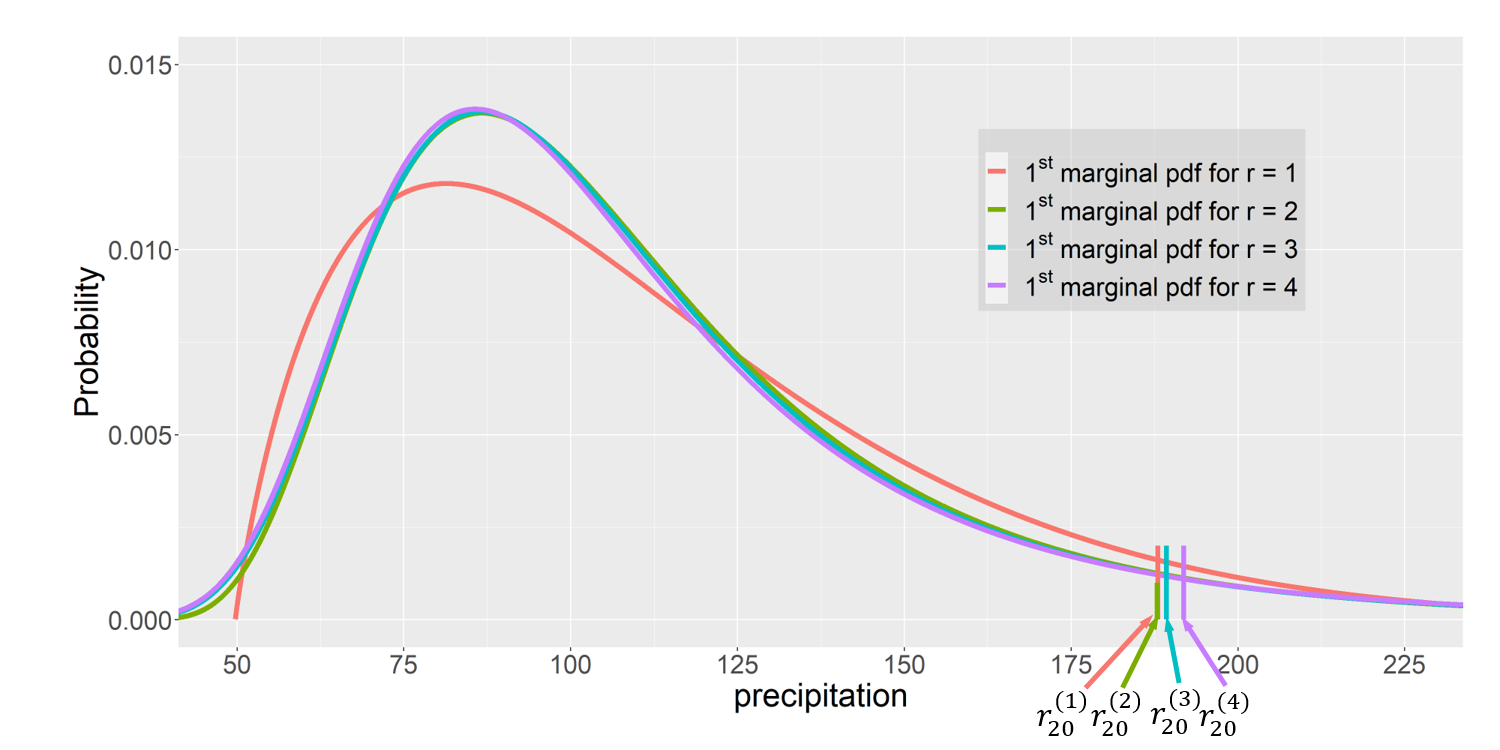} \end{tabular}	
	\caption{Graphs of the first marginal probability density functions (PDF) from the rK4D fit to Bangkok rainfall data with  $r=1,2,3,4$. The 20-year return levels obtained from each PDF are marked at the bottom right with notation $r^{(r)}_{20}$.  } \label{1st-mpdf}
\end{figure}

\begin{figure}[!h]
	\centering
	\begin{tabular}{l}	\includegraphics[width=14.5cm, height=7.5cm]{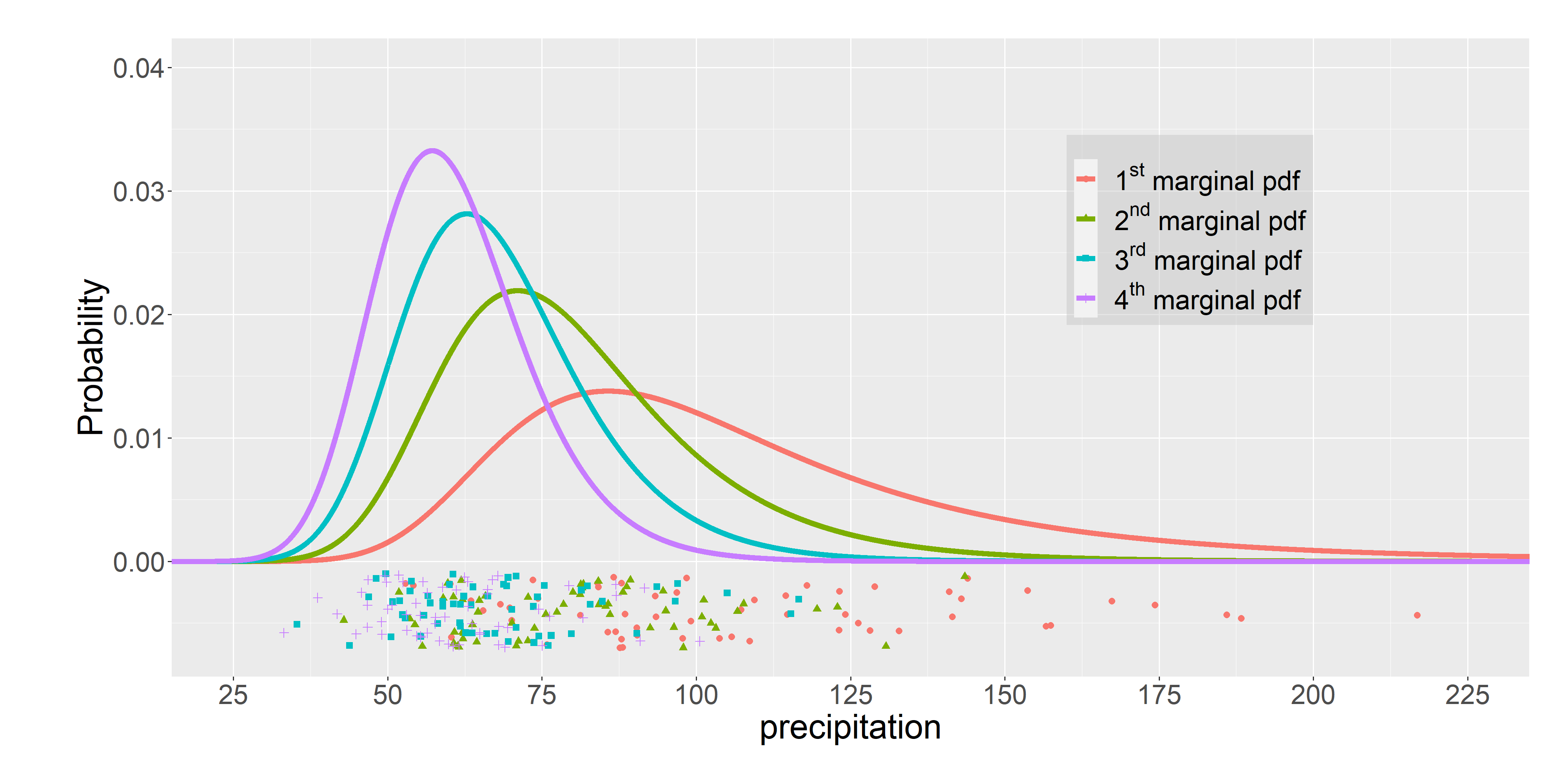} \end{tabular}	
	\caption{Graphs of the $s$-th marginal probability density functions (PDF) from the rK4D/MLE fit to the Bangkok extreme rainfall data with $r=4$, where $s=1,2,3,4$.} \label{1-4mpdf}
\end{figure}

Figure \ref{1st-mpdf} shows graphs of the first marginal PDFs from the rK4D/MLE fit to Bangkok rainfall data with $r=1,2,3,4$. We can conclude from this figure that the first marginal PDFs have a slight change as $r$ increases.
Figure \ref{1-4mpdf} shows graphs of the $s$-th marginal PDFs for $s=1,2,3,4$ from the rK4D/MLE fit with $r=4$. The dot plot of
the $s$-th order statistics are provided at bottom.
We see the shapes of the $s$-th marginal PDFs and how they change as $s$ moves from 1 to 4. By comparing the density curves to the bottom dots, it appears that the $s$-th marginal PDFs fit well with the $s$-th extreme observations.

\section{ {More distributions for $r$-largest order statistics}}

As the K4D includes some distributions as special cases, we can define some  {new distributions for $r$-largest order statistics} as special cases of rK4D.
When $h \rightarrow -1$ in rK4D, the PDF of  {the generalized logistic distribution for $r$-largest order statistics} (rGLD) is obtained as follows:
\begin{linenomath*}\begin{equation} \label{rgld}
\lim_{h\to -1}f_{4}(\underline x^{r})=\sigma^{-r}C_r\times g(\underline x^{r})   
\times\left\{1+\;
 w(x^{(r)})^{1 \over k}\right\}^{-(1+r)} ,
	\end{equation}\end{linenomath*}	
where $C_r (-1)$ is the same as $C_r$ defined in Eq.~(\ref{C_r}) with $h=-1$, $g(\underline x^{r})$ is defined by Eq.~(\ref{g(x)}),  and $w(x)$ is defined by Eq.~(\ref{w(x)}).
This is the $r$-largest extension from a generalized logistic distribution (Ahmad et al.~1988; Hosking and Wallis 1997).

When $h \rightarrow -1$ and $k \rightarrow 0$ in rK4D, the PDF of  {the logistic distribution for $r$-largest order statistics} (rLD) is obtained as a special case of the rGLD as follows:
\begin{linenomath*}\begin{equation} \label{rlogistd}
	\lim_{(h,\:k)\to(-1,\:0)}f_{4}(\underline x^{r})=\sigma^{-r}C_r (-1)
	\times \left\{1+\text{exp}[-\alpha(x^{(r)})]\right\}^{-(1+r)} \times \prod_{j=1}^{r}\text{exp}[-\alpha(x^{(j)})],  	 
	\end{equation}\end{linenomath*}	
where $\alpha(x) = (x-\mu)/\sigma$.

When $k \rightarrow 0$ in rK4D, the PDF of  {the generalized Gumbel distribution for $r$-largest order statistics} (rGGD) is obtained as follows:
\begin{linenomath*}\begin{equation} \label{rggd}
\lim_{k\to 0}f_{4}(\underline x^{r})=\sigma^{-r} C_{r}\times
\left\{1-h\;\text{exp}[-\alpha(x^{(r)})]\right\}^{{1-rh} \over {h}} \times
\prod_{j=1}^{r}\text{exp} [-\alpha(x^{(j)})] .
 \end{equation}\end{linenomath*}
This is the $r$-largest extension from a generalized Gumbel distribution (Jeong et al.~2014).

Because one support condition ($C_r >0 $) of rK4D is $ h < {1 / (r-1)}$ for $r \ge 2$,
rK4D does not include the $r$-largest extensions of the generalized Pareto and exponential distributions (the K4D with $h=1$),
except in the case where $r=1$. Figure \ref{rk4d_rel} shows the relationship between rK4D and the other $r$-largest distributions.
These three parameters for $r$-largest distributions derived from rK4D may serve to flexibly model the $r$-largest extreme values,
with competent performance to the rGEVD.

\begin{figure}[!h]
\centering
 \begin{tabular}{l}	\includegraphics[scale = 0.33]{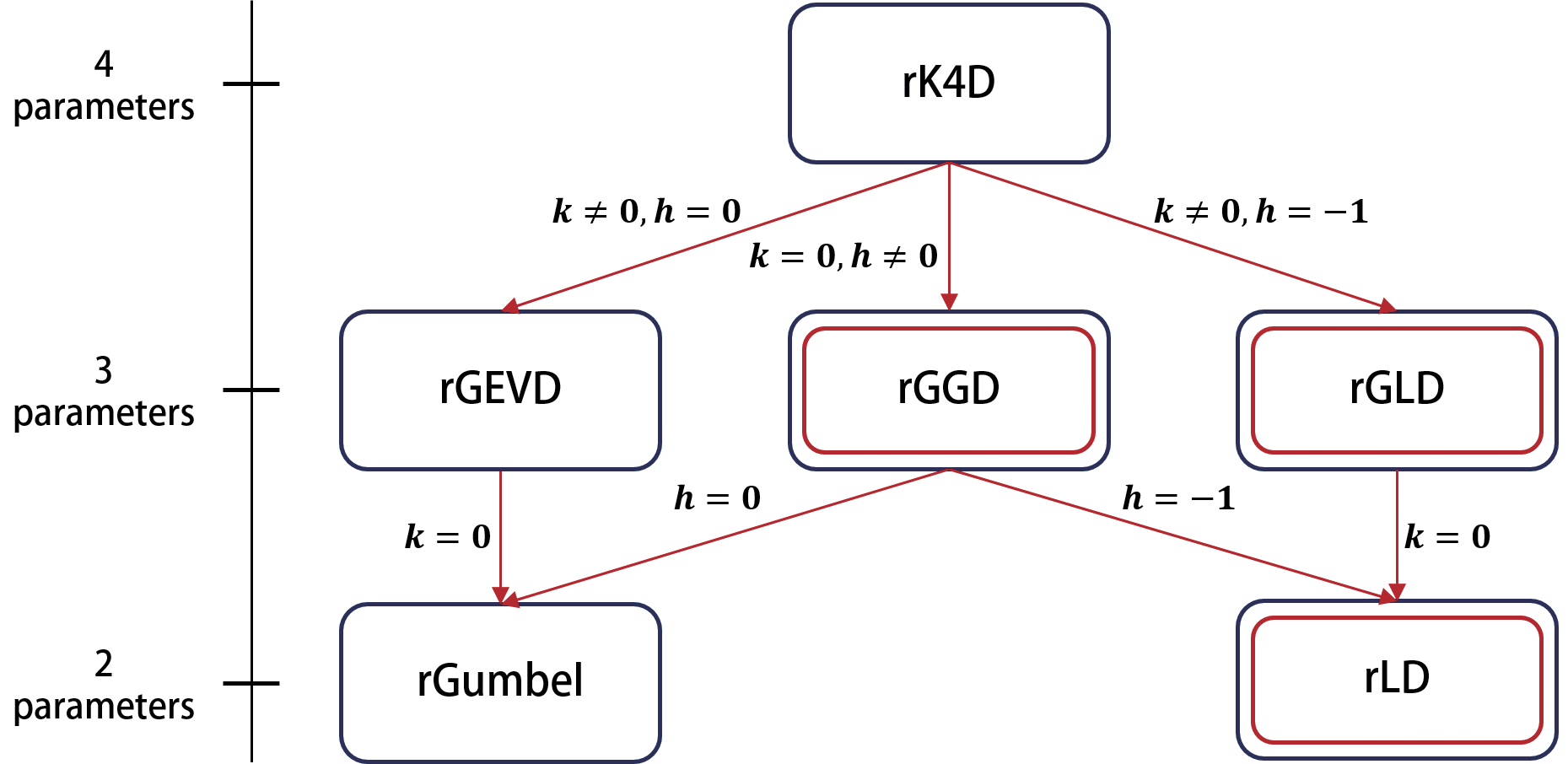} \end{tabular}	
 \caption{Relationship of the  {the four-parameter kappa distribution for $r$-largest order statistics} (rK4D) to other $r$-largest distributions.} \label{rk4d_rel}
\end{figure}

\section{Discussion}

Hosking (1994) derived K4D by applying a transformation $X= \xi +\alpha (1- e^{-k Y}) /k $ to a random variable $Y$, where $Y$ has an exponential, a Gumbel, or a logistic distribution. The rK4D may be obtained similarly by applying a vector transformation to a random vector $\underline{Y}$ of the $r$-largest order statistics. We leave this derivation for future work. A point process approach for extremes (Smith 1989; Coles 2001) may provide a theoretical insight.


The standard error (SE) of the MLE of the rK4D parameters decrease in general as $r$ increase, but the SE of the return level does not show a monotonic reduction trend. When sample size is large such as more than 100, it seems that the variance reduction effect by the addition of the $r$-largest to the first maximum is small, based on our experience. This is maybe because the GEVD or the K4D already estimate the return levels well with large sample, so may not really need to add the $r$-largest observations. We need more study on this matter.  

The use of the $r$-largest extremes enhances the power of the estimation for moderate values of $r$, However, the use of larger values of $r$ may lead to bias in estimation (Zhang et al.~2004). The selection of $r$ is thus important in the rGEVD or in the rK4D. Bader et al.~(2017) and Silva et al.~(2021) developed automated methods for selecting $r$ from the rGEVD. Their approaches (with modifications if necessary) can be applied to rK4D, even though we did not try it in this study. Non-stationary modeling in the rK4D under changing climate is another challenge to be faced in future work.

\section{Conclusion}

In this study, we introduced  {the four-parameter kappa distribution for $r$-largest order statistics} (rK4D) as an extension of  {the generalized extreme value distribution for $r$-largest order statistics} (rGEVD). The joint PDF and marginal and conditional distributions of rK4D were derived. The maximum likelihood and maximum penalized likelihood estimation (MPLE) methods are considered for the parameter estimation of rK4D.
A Monte Carlo simulation study and an application to the Bangkok extreme rainfall data are presented in comparison with the rGEVD. This study concludes that rK4D with MPLE provides better fitting than rGEVD and rK4D with MLE.
Some  {new distributions for $r$-largest order statistics} were also derived as special cases of the rK4D. 

Effective use of the available information is important in extremes because extreme values are scarce. 
Thus, the use of the $r$-largest method is encouraged (Zhang et al.~2004; An and Pandy 2007; Wang and Zhang 2008; Silva 2021). For example, Kim et al.~(2020) used observations and historical data from climate simulation models for the annual maxima of temperature and precipitation during 1981-2000, to fitting the GEVD with the L-moment estimation method (Hosking and Wallis, 1997). For a situation like this case, as the sample size of the block maxima is as small as 20, the rK4D with MPLE approach can result in less bias and uncertainty in estimating the T-year return level of extreme values than the GEVD with the L-moment estimation. 

The rK4D, as an extension of the rGEVD, can serve to
model the $r$-largest observations flexibly, especially when three parameters in the rGEVD are not sufficient to capture the variability of observations well.
We believe that the introduction of rK4D will enrich and improve our modeling methodology for extreme events.

\section*{Acknowledgment}
	The authors would like to thank the reviewer and the Associate Editor for their valuable comments and constructive suggestions.
	This work was supported by the National Research Foundation (NRF) of Korea  (No.2021R1A6A3A13044162, No.2020R1I1A3069260), and BK21 FOUR (No.5120200913674) funded by the Ministry of Education and the NRF of Korea. The authors are grateful to Mr.~Juyoung Hong for his help in writing this article.

\section*{Appendix A.1}
\subsection*{Consistency of the definition of the rK4D}

To check the consistency of the definition Eq. (\ref{rk4d}), we first derive the joint PDF of the $(r-1)$K4D from that of the rK4D by integration.
Then we check whether the PDF of the $(r-1)$K4D still has the same pattern with the PDF (Eq.~(\ref{rk4d})) of the rK4D or not.
If the answer is yes, the definition Eq.~(\ref{rk4d}) is consistent.
\begin{linenomath*}\begin{equation} \label{r-1k4d}   \begin{aligned}
	f_4 (\underline{x}^{r-1}) &= \int_{-\infty}^{x^{(r-1)}} f_4 (\underline{x}^r) dx^{(r)} \\
	 & = \sigma^{-r} C_{r}\times g(\underline x^{r-1})\times\int_{-\infty}^{x^{(r-1)}}  w(x^{(r)})^{{1 \over k} -1}\times\left\{1-h\; w(x^{(r)})^{1 \over k}\right\}^{1-rh \over h}dx^{(r)} \nonumber \\
	     \end{aligned}
	 \end{equation}\end{linenomath*}
 \begin{linenomath*}\begin{equation} 
 \begin{aligned}
	 & substitute \ \ v = 1-h \; w(x^{(r)})^{1 \over k} , \ \ dx^{(r)}=\dfrac{\sigma}{h}\times\ {  w(x^{(r)})^{-{1 \over k} +1}}dv\\
     & = \sigma^{-r}\times C_{r-1}\times g(\underline x^{r-1})\times\dfrac{\sigma}{h}\times\int_{0}^{1-h \; w(x^{(r-1)})^{1 \over k}}v^{1-rh \over h}dv \\
     & =\sigma^{-(r-1)}\times C_{r-1}\times g(\underline{x}^{r-1}) \times F_4 (x^{(r-1)} )^{1-(r-1)h}.
	 \end{aligned}
\end{equation}\end{linenomath*}
\noindent
Now, we see the same pattern  with the PDF (Eq.~(\ref{rk4d})) of the rK4D. Thus our definition (\ref{k4d}) is consistent.

\section*{Appendix A.2}
\subsection*{Margianl probability density functions of the rK4D}

The marginal PDF of $s$-th order statistic from the rK4D is derived to the following
by consecutive integrals of $f_4 (\underline{x}^s)$ with respect to $(x^{(1)}, \dots, x^{(s-1)})$ for $ 2 \le s \le r$.:
\begin{linenomath*}\begin{equation}  \begin{aligned}
    \textcolor{blue}{ p_{4}(x^{(s)}) } &=\int_{x^{(s)}}^{\infty}\int_{x^{(s-1)}}^{\infty}\dots\int_{x^{(2)}}^{\infty}f_{4}(\underline x^{s}) \;dx^{(1)}dx^{(2)} \dots dx^{(s-1)}\\
    & =\sigma^{-s}\: {C_{s}\: F_{4}(x^{(s)})^{1-sh}}\: g(x^{(2)},\dots ,x^{(s)})
    \int_{x^{(s)}}^{\infty}\dots \int_{x^{(2)}}^{\infty} w(x^{(1)})^{{1 \over k} -1} \;dx^{(1)}\dots dx^{(s-1)} \nonumber 
     \end{aligned}
    \end{equation}\end{linenomath*}
    \begin{linenomath*}\begin{equation} \label{margk4d}
    	 \begin{aligned}
    & substitute \ \ v = 1-\dfrac{k(x^{(1)}-\mu)}{\sigma},~~  dx^{(1)}=-\dfrac{\sigma}{k}\;dv  \\
    & =\sigma^{-(s-1)}\: {C_{s}\: F_{4}(x^{(s)})^{1-sh}}\: g(x^{(3)},\dots, x^{(s)})
   \int_{x^{(s)}}^{\infty}\dots \int_{x^{(3)}}^{\infty}{w(x^{(2)})}^{{2 \over k} -1}\;dx^{(2)}\dots dx^{(s-1)}  \nonumber  
      \end{aligned}
   \end{equation}\end{linenomath*}
\begin{linenomath*}\begin{equation} 
\begin{aligned}
    &=\sigma^{-(s-2)}\: {C_{s}\: F_{4}(x^{(s)})^{1-sh}}\: g(x^{(4)},\dots ,x^{(s)})
    \int_{x^{(s)}}^{\infty}\dots \int_{x^{(4)}}^{\infty}\dfrac{1}{2}{w(x^{(3)})}^{{3 \over k} -1}\;dx^{(3)}\dots dx^{(s-1)}  \\
    & ~~~~~~~~~~ \vdots  \\
    & =\sigma^{-1} {{C_s} \over {(s-1)!} }\; w(x^{(s)})^{{s \over k}-1}
	  \times F_4 (x^{(s)} )^{1-sh} , 
      \end{aligned}
\end{equation}\end{linenomath*}		
for $2 \le s \le r$, where $g(\underline{x})$ is defined as in Eq.~(\ref{g(x)}), $w(x)$ is defined as in Eq.~(\ref{w(x)}), and $F_4$ is the CDF of K4D as in Eq.~(\ref{cdf4}). \textcolor{blue}{Here, we use the notation $ p_{4}(x^{(s)})$ to differentiate with $f_4(x)$ in Eq.~(\ref{k4d}). }

We can see, as $h \rightarrow 0$,  that the above marginal PDF in Eq.~(\ref{margk4d}) goes to the corresponding marginal PDF of the rGEVD,
\begin{linenomath*}\begin{equation} \label{marggevd}
\textcolor{blue}{p_3 (x^{(s)}) } =  \sigma^{-1} {1 \over {(s-1)!} }\; w(x^{(s)})^{{s \over k} -1} \times \text{exp} [-\tau(x^{(s)}) ],
\end{equation}\end{linenomath*}	
where
\begin{linenomath*}\begin{equation} \label{tau(x)}
\tau(x^{(s)})= w(x^{(s)})^{1 \over k} = \left[1 -k {{x^{(s)}-\mu} \over {\sigma}} \right]^{1 \over k}.
\end{equation}\end{linenomath*}	
This marginal PDF of the rGEVD is also obtained by differentiating the marginal CDF of the rGEVD which is:
\begin{linenomath*}\begin{equation} \label{margcdfgevd}
H_3( x^{(s)}) = \text{exp} [-\tau(x^{(s)}) ]\; \sum_{i=0}^{s-1} { {\tau(x^{(s)})^i} \over {i!}  },
\end{equation}\end{linenomath*}	
as provided in Coles (2001, p.67). 


\section*{Appendix A.3}
\subsection*{Margianl distribution functions of the rK4D}

The marginal CDF of $s$-th order statistic from the rK4D is \textcolor{blue}{ obtained by integrating $p_4 (x^{(s)})$ as follows: for $h>0$,}
\begin{linenomath*}\begin{equation} \label{margcdfk4d} 
\begin{aligned}
	P(x^{(s)}<t) &\triangleq H_4( x^{(s)}) = \int_{-\infty}^t \textcolor{blue}{p_4 (x^{(s)})} dx^{(s)} \\
&=\int_{-\infty}^{t}\sigma^{-1}\times\dfrac{C_{s}}{(s-1)!}\times[1-\dfrac{k(x^{(s)}-\mu)}{\sigma}]^{{s \over k} -1}\times\left\{1-h[1-\dfrac{k(x^{(s)}-\mu)}{\sigma}]^{1 \over k}\right\}^{1-sh \over h}dx^{(s)} \nonumber
	   \end{aligned}
\end{equation}\end{linenomath*}
\begin{linenomath*}\begin{equation} 
\begin{aligned}
	&substitute \ v= 1-\dfrac{k(x^{(s)}-\mu)}{\sigma},\: dx^{(s)}= -\dfrac{\sigma}{k}dv\\
	&=\dfrac{C_{s}}{(s-1)!}\times{1 \over k}\times\int_{1-{{k(t-\mu)} \over {\sigma}}}^{\infty}v^{{s \over k} -1}\times[1-hv^{1 \over k}]^{{1-sh} \over {h}}dv \nonumber 
	   \end{aligned}
	\end{equation}\end{linenomath*}
\begin{linenomath*}\begin{equation} 
\begin{aligned}
	&substitute \ w= 1-hv^{1 \over k},\: dv= -{{k} \over {h}}v^{-({1 \over k} -1)}dw\\
	&=\dfrac{C_{s}}{(s-1)!}\times{1 \over {h^{s}}}\times\int_{0}^{{1-h[1-{{k(t-\mu)} \over {\sigma}}]}^{1 \over k}}w^{{1-sh} \over {h}}(1-w)^{s-1}dw\\
	&=\dfrac{C_{s}}{(s-1)!}\times{1 \over {h^{s}}}\times \textcolor{blue}{ IB\left( F_4^h (t); \: a_s,\: s \right) } \\
	&\ \textcolor{blue}{ = \dfrac{\ IB\left( F^h_4 (t); \: a_s,\: s \right)}{ B (a_s, \: s)}, }
\end{aligned}
	\end{equation}\end{linenomath*}		
\textcolor{blue}{ where $a_s = \{1-(s-1)h\}/h $, $\ B(u,v)$ is a beta function,}
	 and $IB(x; u,v)$ is the incomplete beta function defined as follows:
\begin{linenomath*}\begin{equation} \label{IB} \begin{aligned}
	IB(x; u,v) &= \int_{0}^x t^{u-1}(1-t)^{v-1}dt, \quad u, v>0; \ 0<x<1. 
	\end{aligned}
	\end{equation}\end{linenomath*}	
\textcolor{blue}{ The last equality in Eq.~(\ref{margcdfk4d}) is obtained from that }
\begin{linenomath*}\begin{equation} 
\textcolor{blue}{\dfrac{C_{s}}{(s-1)!}\; \left({1 \over {h}}\right)^s \times B(a_s,\: s )  = 1~~ \text{for}~~ h>0. \nonumber }
	\end{equation}\end{linenomath*}	

\textcolor{blue}{When $h<0$, by letting $U = F_4 ^{(-h)} (x^{(s)})$ and using }
\begin{linenomath*}\begin{equation}
\textcolor{blue}{f_4 (x^{(s)}) = \sigma^{-1}\; ( 1/h )^{1-k}\;\{1- F_4^h (x^{(s)}) \}^{1-k}\; F_4^{1-h} (x^{(s)} ),  } \nonumber 
	\end{equation}\end{linenomath*}	
\textcolor{blue}{we have }
	\begin{linenomath*}\begin{equation}
			\begin{aligned}
H_4(t)	& \textcolor{blue}{= \dfrac{C_{s}}{(s-1)!} \ \left( {1 \over {-h}} \right)^{s} \times  IB \left( F_4^{(-h)} (t); 
\: {{1} \over {-h}},\: s \right),    }\nonumber \\
& \textcolor{blue}{=  \dfrac{\ IB \left( F_4^{(-h)} (t); \: {{1} \over {-h}},\: s \right)}{ B \left( {1 \over {-h}},\: s \right) } .     } \nonumber 
	\end{aligned}
	\end{equation}\end{linenomath*}	
\textcolor{blue}{because}
	\begin{linenomath*}\begin{equation}
\textcolor{blue}{\dfrac{C_{s}}{(s-1)!}\; \left({1 \over {-h}}\right)^s \times B \left( {1\over {-h}},\: s \right) = 1 ~~ \text{for}~~ h<0 . }  \nonumber 
		\end{equation}\end{linenomath*}	
\textcolor{blue}{Thus, this marginal CDF of $s$-th order statistic from the rK4D, $H_4( x^{(s)})$ is actually the CDF of a beta distribution.}
When $h \rightarrow 0$, this marginal CDF goes to the marginal CDF of the rGEVD in Eq.~(\ref{margcdfgevd}).

The quantiles from this marginal CDF are obtained by a constant times the quantiles of the beta distribution, because the function $IB(x; u,v)/B(u,v)$ is actually the CDF of a beta distribution.


\section*{Appendix A.4}
\subsection*{Conditional distributions of the rK4D}

The conditional PDF of $X^{(s)}$ given $\underline{X}^{s-1}$ is, for $2 \le s \le r$:
\begin{linenomath*}\begin{equation} \label{condpdfk4d} \begin{aligned}
f_4 (x^{(s)} | \underline{x}^{s-1}) &= {f_{4}(\underline{x}^{s}) \over f_{4}(\underline{x}^{s-1})} \\
   &=  \frac{ \sigma^{-1} \times (1-(s-1)h) \times g(x^{(s)}) \times F_4 (x^{(s)})^{1-sh}}{ F_4(x^{(s-1)})^{1-(s-1)h} } , \\
   &\ \textcolor{blue}{ =  \frac{ \{1-(s-1)h\} \; f_4 (x^{(s)}) \;  F_4 (x^{(s)})^{-(s-1)h} } {F_4 (x^{(s-1)})^{1-(s-1)h } }, \nonumber }
   \end{aligned}
 	\end{equation}\end{linenomath*}	
 under $ x^{(s)} \le x^{(s-1)}$, \textcolor{blue}{where $g(x^{(s)}) =  w(x^{(s)})^{{1 \over k} -1}$
and $w(x)$ is defined in Eq.~(\ref{w(x)}); $ w(x^{(s)})= 1 -k {{x^{(s)}-\mu} \over {\sigma}}$.}
 The Markov property is thus satisfied.

 The conditional CDF of $X^{(s)}$ given $\underline{X}^{s-1}$ is,
 \begin{linenomath*}\begin{equation} \label{condcdfk4d}
 	H_4 (x^{(s)} | \underline{x}^{s-1}) =   \frac{F_4 (x^{(s)})^{1-(s-1)h} }{ F_4
 		(x^{(s-1)})^{1-(s-1)h} } = \left(\frac{F_4 (x^{(s)})} { F_4
 		(x^{(s-1)})}\right)^{1-(s-1)h} ,
 	\end{equation}\end{linenomath*}	
 under $ x^{(s)} \le x^{(s-1)}$.
 This is same as  \textcolor{blue}{the $({1-(s-1)h})$ power of} the CDF of the K4D with right truncated at $x^{(s-1)}$ (Johnson et al.~1995). This property is used in
 generating random numbers from the rK4D.
 
 \section*{Appendix A.5} 
 \subsection*{Random number generation of rK4D samples}
 
\textcolor{blue}{ The property in Eq.~(\ref{condcdfk4d}) 
 is exploited to generate $r$ components in a realized rK4D observation, by modifying the algorithm of Bader et al.~(2017) for the rGEV. A pseudo algorithm follows: }
 \begin{itemize} 
 	\item[1.] Generate $U_1, \dots, U_r$, where $U$ values are random numbers from the uniform distribution (0,1), {for a fixed $r$}.
 \textcolor{blue}{	\item[2.] Truncate uniform random numbers ($W_1, \dots, W_r$) as $\ W_s = \prod_{j=1}^s\ (U_j)^{b_j} \ $ for $s=1,\dots,r$, where $\ b_j = 1 / \{1-(j-1)h\}$.}
 	\item[3.] Obtain $x^{(i)} = F^{-1}_4 (W_i)$, where $F_4$ is the CDF of the K4D as given in Eq.~(\ref{cdf4}).
 \end{itemize}
 The resulting vector $(x^{(1)}, \dots, x^{(r)})$ is a single observation from the rK4D.

\section*{Appendix B.1}
\subsection*{Quantiles of the $s$-th order statistic of the GEVD and of the K4D}

The quantiles of the GEVD are obtained by inverting Eq. (\ref{cdf-gevd}):
\begin{linenomath*}\begin{equation}\label{z_pgevd}
z_p = \mu + {\sigma \over k } [ 1- \{ - \text{log}(1-p) \} ^{k} ],
\end{equation}\end{linenomath*}	
where $F_3 ( z_p )= 1-p $. Here, $z_p$ is known as the return level associated with
the return period $1/p$, since the level $z_p$ is expected to be exceeded on
average once every $1/p$ years (Coles 2001). For example, a 20-year (50-year)
return level is computed as the 95th (98th) quantile of the fitted GEVD.

The quantiles of the K4D are
\begin{linenomath*}\begin{equation} \label{z_pk4d}
z_p= \mu +\dfrac{\sigma}{k} \left\{ 1-\left(\dfrac{1-(1-p)^h}{h}\right)^k \right\},
\end{equation}\end{linenomath*}	
where $F_4 ( z_p )= 1-p $.
The quantiles of the $s$-th order statistic from the rK4D are obtained by inverting the $s$-th marginal CDF in {the Appendix A.3.} 

\section*{Appendix B.2}
\subsection*{Delta method for variance estimation in the K4D} 

The variance estimation of the $1/p$-year return level ($z_p$) can be calculated by the delta method (Coles, 2001).
We present the details of the procedure including the derivatives of $z_p$ with respect to each parameter as follows;
\begin{linenomath*}\begin{equation} \label{delta}
	\text{Var} (\hat z_p) \approx \nabla z_p^t \;V \;\nabla z_p ,
	\end{equation}\end{linenomath*}	
where $V$ is the covariance matrix of parameter estimates which is approximated by the inverse of the observed Fisher information matrix, and
\begin{linenomath*}\begin{equation}\label{z_pgevd1}
\nabla z_p^t =[\dfrac{\partial z_{p}}{\partial\mu},\:\dfrac{\partial z_{p}}{\partial\sigma},\:\dfrac{\partial z_{p}}{\partial k},\:\dfrac{\partial z_{p}}{\partial h}].
\end{equation}\end{linenomath*}
The components in Eq.~(\ref{z_pgevd1}) are\\
\begin{linenomath*}\begin{equation}\label{z_pgevd2}
\dfrac{\partial z_{p}}{\partial\mu}=1, ~~~
\dfrac{\partial z_{p}}{\partial\sigma}=\dfrac{1-y_p^{k}}{k},
\end{equation}\end{linenomath*}
\begin{linenomath*}\begin{equation}\label{z_pgevd4}
\dfrac{\partial z_{p}}{\partial k}= -\dfrac{\sigma\ln(y_p) \times y_p^{k}}{k}-
\dfrac{\sigma(1-y_p^{k})}{k^{2}},
\end{equation}\end{linenomath*}
\begin{linenomath*}\begin{equation}\label{z_pgevd5}
\dfrac{\partial z_{p}}{\partial h}=\dfrac{\sigma\left\{(1-p)^{h}h \;\ln(1-p) \;+1-(1-p)^{h}\right\}}{h^{2}} \times y_p^{k-1},
\end{equation}\end{linenomath*}
where $y_p=\dfrac{1-(1-p)^{h}}{h}$, evaluated at $(\hat \mu,\; \hat \sigma,\; \hat k,\; \hat h)$. 

\section*{Appendix B.3}
\subsection*{Profile likelihood for confidence interval of the return level in the rK4D}

For the confidence interval of the return level, the profile likelihood approach (Coles, 2001) can be useful. This requires a reparameterization of the rK4D model, so that $z_p$ is one of the model parameters, after which the profile log-likelihood is obtained by maximization with respect to the remaining parameters in the usual way. Reparameterization is straightforward:
\begin{linenomath*}\begin{equation} \label{profile}
\mu= z_p-\dfrac{\sigma}{k} \left\{ 1-\left(\dfrac{1-(1-p)^h}{h}\right)^k \right\},
\end{equation}\end{linenomath*}	
so that replacement of $\mu$ in Eq.~(\ref{lfrk4d}) with Eq.~(\ref{z_pk4d}) has the desired effect of expressing the rK4D model in terms of the parameters ($z_p$, $\sigma$, $k$, $h$). The $95\%$ confidence interval for $z_p$ is obtained by calculating a value at a height of 0.5 $\times$ $C_{1,0.05}$ below the maximization of profile log-likelihood, where $C_{1,0.05}$ is the 95\% quantile of a $\chi_1^2$ distribution, and reading off the points of intersection (Coles, 2001).

\section*{Appendix C: Supplementary data}

Supplementary material related to this article can be found online at 
doi.org/10.1016/j.wace.2022.100533. We provide more details
on accurate computation, penalty functions, power evaluation, rainfall
data, and simulation result. 

\renewcommand{\baselinestretch}{0.8}


\end{document}